\documentclass[aps,article,author-year,notitlepage,showpacs]{revtex4-1}
\usepackage{subeqn}
\usepackage{graphicx}
\usepackage{float}
\usepackage[T1]{fontenc}
\usepackage[latin1]{inputenc}
\usepackage{amssymb}
\usepackage{subfigure}
\include{epsf}
\epsfverbosetrue

\newcommand{\bea}{\begin{eqnarray}}
\newcommand{\eea}{\end{eqnarray}}

\makeatother
\begin{document}
\title{4-point vertices from the 2PI and 4PI Effective Actions}

\author{M.E. Carrington}
\email[]{carrington@brandonu.ca} \affiliation{Department of Physics, Brandon University, Brandon, Manitoba, R7A 6A9 Canada}\affiliation{Winnipeg Institute for Theoretical Physics, Winnipeg, Manitoba}

\author{Wei-Jie Fu}
\email[]{fuw@brandonu.ca} \affiliation{Department of Physics, Brandon University, Brandon, Manitoba, R7A 6A9 Canada}
\affiliation{Winnipeg Institute for Theoretical Physics, Winnipeg,
Manitoba}

\author{P. Mikula}
\email[]{pnmikula@gmail.com} \affiliation{Department of Physics, University of Manitoba, Winnpieg, Manitoba, R7A 6A9 Canada}
\affiliation{Winnipeg Institute for Theoretical Physics, Winnipeg,
Manitoba}

\author{D. Pickering}
\email[]{fuw@brandonu.ca} \affiliation{Department of Mathematics, Brandon University, Brandon, Manitoba, R7A 6A9 Canada}

\date{\today}

\begin{abstract}
We consider a symmetric scalar theory with quartic coupling in 2-
and 3-dimensions and compare the self-consistent 4-point vertex obtained from the 
4PI effective action with the Bethe-Salpeter 4-vertex from 2PI effective
action. At zero external momenta the two vertices agree well with each other when the coupling strength is small, but differences
between them become more and more pronounced as the coupling
strength is increased. We also study the momentum dependence of the two
vertices and show that for certain momentum
configurations they are almost identical, but differ for general
momentum arguments.
\end{abstract}


\large
\maketitle

\section{Introduction}
\vspace{5pt}

The resummation of certain classes of Feynman diagrams to infinite loop order
is a useful method in quantum field theory when there is no small expansion parameter and standard perturbative methods do not apply.
A familiar example
is the hard thermal loop theory~\cite{Braaten1990} which was developed in the context of thermal field
theory.
The $n$PI effective action formalism \cite{Luttinger1960,Cornwall1974,Norton1975,Carrington2004,Berges2004,Carrington2010,Carrington2011} is another approach in which the action is expressed as a functional of non-perturbative vertex functions which are determined through
a self-consistent stationary equation after the effective action is
expanded to some order in the loop or $1/N$ expansion.
These self-consistent equations of motion resum certain classes of diagrams to infinite order, and the classes that are resummed are determined by the set of skeleton diagrams that are included in the truncated effective action.

The 2PI effective action formalism has been used to
describe equilibrium thermodynamics  \cite{Blaizot1999,Berges2005a}
and the
quantum dynamics of far from equilibrium of quantum fields (see \cite{Berges2001} and references
therein).
One important application is the calculation of transport coefficients \cite{Aarts2004,
Carrington2006}.

The 2PI theory does not contain a self-consistent 4-point vertex, but one can 
obtain a non-perturbative 4-point function through the Beta-Salpeter
(BS) equation. 
It is of interest to compare this 4-vertex with the self-consistent 4-vertex from the 4PI effective action. 
In this paper we compare numerically the BS 4-vertex (M) obtained from the 2PI effective action and the self-consistent 4-vertex (V) from the 4PI effective action. 
In general the two vertices are quite different, but if we compare the BS vertex with the 4PI vertex in a specific momentum configuration $V(P,-P,K,-K)$ (which we call the ``diagonal'' configuration), we find good agreement between them. The conclusion is that for
physical quantities for which the physics of the 4-point function is
important, but for which diagonal momenta are expected to dominate,
accurate results can be expected using the 2PI effective action. In
general however, one needs the full non-perturbative 4-point
function obtained from the 4PI effective action. An example is the
shear viscosity which can be obtained in the large $N$ approximation
from the 2PI effective theory, but at leading order one must use a
higher order effective action \cite{Carrington2008,Carrington2009,Carrington2010b}.

Numerical calculations in higher order $n$PI theories are extremely difficult and little progress has been made. One of the main obstacles is the size of the phase space involved. In this paper we develop a technique to reduce the region of the phase space that is needed to calculate $V$ by maximally exploiting the symmetries of the vertex. This technique allows us to extend results reported in a previous paper \cite{Carrington2013} to larger lattice sizes.  

This paper is organized as follows. In section \ref{generalSection}
we review the $n$PI formalism and define our notation. In
sections \ref{numericalSect} we describe our numerical technique, and in \ref{ResultSect} we present 
results from numerical calculations in 2D and 3D. In section
\ref{concSect} we give our conclusions.

\section{General Formalism}
\label{generalSection}
\vspace{5pt}

We consider the following Lagrangian:
\begin{equation}
\label{lagrangian}
{\mathcal{L}}=\frac{1}{2}(\partial_{\mu}\varphi\partial^{\mu}\varphi-m_b^{2}{\varphi}^{2})
-\frac{i\lambda_b}{4!}\varphi^{4}\,.
\end{equation}
In this equation we have introduced a scaled version of the usual coupling constant defined as $\lambda_b = -i\lambda_{\rm phys}$ for future notational convenience. 
The classical action is:
\bea
\label{defnClass}
&& S[\phi]=S_{0}[\varphi]+S_{\mathrm{int}}[\varphi]\,,\\
&& S_{0}[\varphi]=\frac{1}{2}\int d^d x\, d^dy\,\varphi(x)\big[i G_0^{-1}(x-y)\big]\varphi(y) \,,\nonumber\\
&& S_{\mathrm{int}}[\varphi]=- \frac{i\lambda_b}{4!}\int d^d x  \varphi^4 (x)\,,\nonumber
\eea
where $d$ denotes the number of dimensions. 
We use a single index to represent space-time arguments and use the summation convention to imply integration. 
In many equations in this paper, we suppress these indices altogether. As an example of this notation, the non-interacting part of the classical action is written:
\bea
\label{notex}
&& S_{0}[\varphi]=\frac{1}{2}\int d^d x\, d^dy\,\varphi(x)\big[i G_0^{-1}(x-y)\big]\varphi(y) = \frac{i}{2}\varphi_i G_{0,ij}^{-1}\varphi_j \,,\\
 ~~ \rightarrow ~~&& S_{0}[\varphi] =  \frac{i}{2}G_0^{-1}\varphi^2\,,~~~G_0^{-1}=-i \frac{\delta^2 S_{cl}[\varphi]}{\delta \varphi^2}\bigg|_{\varphi=0}=i(\Box+m_b^2)\,.\nonumber
\eea

\subsection{4PI effective action}

The 4PI effective action is obtained from the Legendre transformation of the connected generating functional:
\bea
\label{genericGamma}
&& Z[R_1,R_2,R_3,R_4]=\int [d\varphi]  \;{\rm Exp}[i\,(S_{cl}[\varphi]+R_1\varphi + \frac{1}{2} R_2\varphi^2 + \frac{1}{3!} R_3\varphi^3 +\frac{1}{4!} R_4\varphi^4)]\,,\\[1mm]
&&W[R_1,R_2,R_3,R_4]=-i \,{\rm Ln} Z[R_1,R_2,R_3,R_4]\,,\nonumber\\[1mm]
&&\Gamma[\phi,G,V_3,V_4] = W - R_1\frac{\delta W}{\delta R_1} - R_2\frac{\delta W}{\delta R_2} - R_3\frac{\delta W}{\delta R_3} - R_4\frac{\delta W}{\delta R_4} \,.\nonumber
\eea
For future use we note the relations:
\bea\label{defcon}
\frac{\delta  W }{\delta R_{1}} &&= \langle\varphi_i\rangle = \phi \,,\\
2\frac{\delta  W }{\delta R_{2}} &&= \langle\varphi
\,\varphi\rangle = G+\phi \phi \,. \eea The Legendre
transforms can be done using the method of subsequent
transformations \cite{Carrington2004,Berges2004} which involves
starting from an expression for the  2PI effective action and
exploiting the fact that the source terms $R_3$ and $R_4$ can be
combined with the corresponding bare vertex by defining a modified
interaction vertex. We consider only the symmetric theory for which
odd $n$-point functions are zero, and from now on we drop the
subscript on the 4-point vertex function and write $V:=V_4$. Writing
the result in terms of renormalized quantities the 4-Loop 4PI
effective action has the form: 
\bea \label{gammaDef}
 i\Gamma[G,V]&&=- \frac{1}{2}\mathrm{Tr}\ln G^{-1} - \frac{1}{2}\mathrm{Tr}G^{-1}_{0} G + \Phi_0[G,V] + \Phi_{\mathrm{int}}[G,V]\,,\\
\label{phiDef}
\Phi_0 &&= - \frac{1}{2}\mathrm{Tr}\delta G^{-1}_{0}G+
\frac{1}{8}(\lambda+\delta\lambda) G^2  +
\frac{1}{4!}(\lambda+\delta\lambda) G^4 V\,. \eea 
We use
\bea \label{phi-defns} i \Gamma =
\Phi=\Phi_1+\Phi_2=\Phi_1+(\Phi_0+\Phi_{\rm int})\,, \eea where

$\Phi_1$ is all terms with less than 2-loops and is given by the first two terms in (\ref{gammaDef})

$\Phi_2$ is all terms with 2 or more loops

$\Phi_0$ is the diagrams in $\Phi_2$ which have either counter-terms
or bare vertices 

$\Phi_{\rm int}$ is the remaining terms in $\Phi_2$ (all of which have 3
or more loops)

\begin{figure}[H]
\begin{center}
\includegraphics[width=16cm]{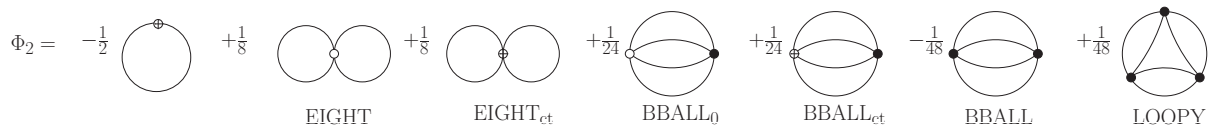}
\end{center}
\caption{$\Phi_2 = \Phi_0+\Phi_{\rm int}$ for the 4-Loop 4PI effective action.
The open circle is a bare vertex, the circle with a cross denotes a
counter-term, and the vertex $V$ is indicated by a solid dot. The first five graphs give $\Phi_0$ and the remaining three are $\Phi_{\rm int}$ to 4-loops. 
\label{PhiAandB}  }
\end{figure}

The functional $\Phi_0$ has the same form for $n\ge
4$ and all orders in the loop expansion, while $\Phi_{\mathrm{int}}$ contains a set of skeleton
diagrams which are determined by the order of the loop expansion.
The sum of these two pieces is shown in figure
\ref{PhiAandB}.\footnote{All figures are drawn using jaxodraw
\cite{jaxo}.} In all diagrams, bare 4-vertices are
represented by white circles, counter-terms are circles with crosses
in them, and solid dots are the vertex $V$. 

\begin{figure}[H]
\begin{center}
\includegraphics[width=14cm]{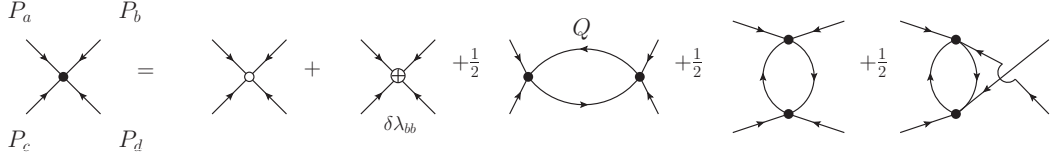}
\end{center}
\caption{Self-consistent equation of motion for the proper
4-point vertex.\label{f4}}
\end{figure}

The self-consistent 2- and 4-point functions in the symmetric phase are obtained by solving simultaneously the equations of motion:
\bea
\label{equation:eom}
\frac{\delta \Gamma[G,V]}{\delta G}\bigg|_{G=\tilde G,V=\tilde V}=0\,,~~~\frac{\delta \Gamma[G,V]}{\delta V}\bigg|_{G=\tilde G,V=\tilde V}=0\,.
\eea
The equation of motion for $V$ is given in figure \ref{f4} and equation (\ref{Eq15}). We use throughout the notation $dQ=d^d q/(2\pi)^d$.
\par
\bea V(P_{a},P_{b},P_{c})&&
=\lambda+\delta\lambda +V_{s}(P_{a},P_{b},P_{c})
+V_{t}(P_{a},P_{b},P_{c})+V_{u}(P_{a},P_{b},P_{c}) \label{Eq15}\,,\nonumber\\[4mm]
V_{s}(P_{a},P_{b},P_{c})&&=\frac{1}{2}\int dQ\; V(P_{a},P_{c},Q)G(Q)G(Q+P_{a}+P_{c})V(P_{b},P_{d},-Q)\,, \\
V_{t}(P_{a},P_{b},P_{c})&&=V_{s}(P_{a},P_{c},P_{b})\,,\nonumber\\[2mm]
V_{u}(P_{a},P_{b},P_{c})&&=V_{s}(P_{a},P_{b},P_{d})\,,~~~P_{d}=-(P_{a}+P_{b}+P_{c})\,.\nonumber
\eea

\begin{figure}[H]
\begin{center}
\includegraphics[width=16cm]{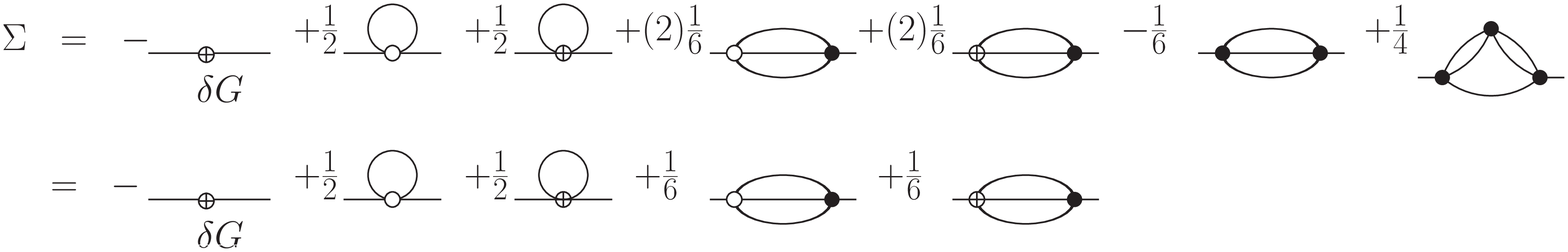}
\end{center}
\caption{The self-energy obtained from equations (\ref{gammaDef}) and (\ref{G2defn}).}\label{SEall}
\end{figure}

The equation of motion for the 2-point function (obtained from the variational equation $\delta\Gamma/\delta G=0$) has the form of a Dyson
equation where the self-energy is proportional to the function derivative of the terms in the effective action with two and more loops: \bea
\label{G2defn} &&G^{-1}=G_0^{-1}-\Sigma\,,~~\Sigma =
2\frac{\delta\Phi_2}{\delta G} \,. \eea
The result is shown in the first line of figure \ref{SEall}. The diagrams can be rearranged by substituting the $V$ equation of motion into the vertex on the left side of the sixth diagram. 
This substitution cancels the 3-loop diagram and produces the result in the second line of the figure and equation (\ref{Eq19}).
\par
\begin{eqnarray}
\Sigma(P) &=&i(\delta Z P^{2}-\delta
m^{2})+\frac{1}{2}(\lambda+\delta\lambda)\int dQ\; G(Q)\nonumber \\
&&+\,\frac{1}{6}(\lambda+\delta\lambda_{\rm bb})\int dQ \int dK\;
V(P,Q,K)G(Q)G(K)G(Q+K+P)\,.\label{Eq19}
\end{eqnarray}

\subsection{2PI Bethe-Salpeter Equation}
\label{section:standardBS}

Although the 2PI effective
action does not explicitly contain a 4-vertex, it is well known that
it can be used to obtain a non-perturbative 4-point vertex called
the Bethe-Salpeter vertex \cite{vanHees2002}. The equation that
determines the 2PI BS vertex is obtained by calculating the
functional derivatives of the 2PI effective action with respect to the
2-point function $G_{kl}$ and the bilocal source. In order to avoid unnecessary subscripts, we use $J$ and
$R$ to represent the external sources $R_{1}$ and $R_{2}$ 
in the 2PI effective action.

\begin{figure}[H]
\begin{center}
\includegraphics[scale=.6]{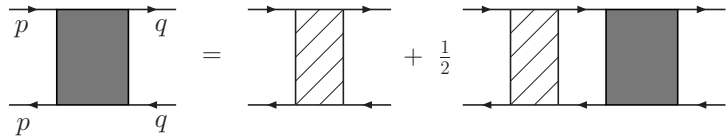}
\caption{Diagrammatic representation of the BS equation in equation
(\ref{BSfirst-mom}). Grey boxes and boxes with oblique lines in them 
represent the vertex $M$ and kernel $\Lambda$, respectively.
}\label{fig:standardBS}
\end{center}
\end{figure}

Using Eqs. (\ref{genericGamma}) and (\ref{phi-defns}) we have directly
\bea
\label{side1}
\frac{\delta}{\delta R_{ij}} \frac{\delta}{\delta G_{kl}}\Phi = -\frac{i}{4}\big(\delta_{ik}\delta_{jl}+\delta_{il}\delta_{jk}\big)\,.
\eea
The derivatives on the left side of (\ref{side1}) can also be written
\bea
\label{side2}
\frac{\delta}{\delta R_{ij}} \frac{\delta}{\delta G_{kl}}\Phi = \frac{\delta \phi_x}{\delta R_{ij}}\frac{\delta^2\Phi}{\delta\phi_x\delta G_{kl}}  +  \frac{\delta G_{xy}}{\delta R_{ij}}\frac{\delta^2\Phi}{\delta G_{xy}\delta G_{kl}}\,.
\eea
Since we consider only the symmetric theory, we drop all terms that correspond to vertices with an odd number of legs, which means that only the second term on the right side survives. Using (\ref{phi-defns}) we write:
\bea
\label{LAM020}
4\frac{\delta^2\Phi}{\delta G_{xy}\delta G_{kl}} && =4\frac{\delta^2\Phi_1}{\delta G_{xy}\delta G_{kl}}+4\frac{\delta^2\Phi_2}{\delta G_{xy}\delta G_{kl}}\,,\nonumber\\[2mm]
&& =: \Lambda^{disco}_{xykl}+ \Lambda_{xykl} = -(G^{-1}_{xk}G^{-1}_{yl}+G^{-1}_{xl}G^{-1}_{yk})+\Lambda_{xykl}\,.
\eea
The term $\Lambda^{disco}_{xykl}$ represents all disconnected contributions and comes from the 1-loop terms in the effective action, and $\Lambda_{xykl}$ contains all contributions from $\Phi_2$.
The derivative of the propagator with respect to the source $R$ is:
\bea
\label{GderR2}
\frac{\delta G_{xy}}{\delta R_{ij}} &&= \frac{\delta}{\delta R_{ij}}\big(\langle \varphi_x\varphi_y\rangle - \langle\varphi_x\rangle\langle\varphi_y\rangle\big)\,, \nonumber\\
&& = \frac{i}{2}\big(\langle \varphi_x \varphi_y \varphi_i \varphi_j\rangle - \langle \varphi_x\varphi_y\rangle  \langle \varphi_i\varphi_j\rangle  ~ + ~ \cdots\big)\,,  \nonumber\\
&& = \frac{i}{2}\big(G_{ia}G_{jb}G_{xc}G_{yd}M_{abcd}+G_{ix}G_{jy}
+G_{iy}G_{jx}~+~\cdots \big)\,, \eea where the dots indicate
expectation values which contain an odd number of field operators
and are zero in the symmetric theory. 
Substituting equations (\ref{LAM020}) and (\ref{GderR2})
into (\ref{side2}) and setting the result equal to the expression
obtained in (\ref{side1}) we have: \bea 0=\frac{i}{4} G_{ix}
G_{jy}\,\big(-M_{klxy}+\Lambda_{klxy}+\frac{1}{2}
\Lambda_{klab}G_{ac}G_{bd} M_{cdxy}\big)\,, \eea where we have used
the fact that the vertices $\Lambda_{klxy}$ and $M_{klxy}$ are
symmetric with respect to permutations of the first two indices, or
the second two indices, or the interchange of the first pair and the
second pair:
$\Lambda_{klxy}=\Lambda_{lkxy}=\Lambda_{klyx}=\Lambda_{xykl}$.
Truncating the external legs we obtain the standard form of the BS
equation: \bea \label{BSfirst-coord}
M_{xykl}=\Lambda_{xykl}+\frac{1}{2} \Lambda_{xyab}G_{ac}G_{bd}
M_{cdkl}\,. \eea We consider systems in thermal equilibrium for
which the system is invariant under space-time translations and write equation 
(\ref{BSfirst-coord}) in momentum space
as: 
\bea 
\!\!\!\!\!\!\!\!\label{BSfirst-mom} M(P,-P,Q,-Q) =
\Lambda(P,-P,Q,-Q)+\frac{1}{2}\int
dK\,\Lambda(P,-P,K,-K)G^2(K)M(K,-K,Q,-Q)\,. \eea 
We will refer to
the momentum arguments in Eq. (\ref{BSfirst-mom}) as ``diagonal.''
Equation (\ref{BSfirst-mom}) is shown diagrammatically in figure
\ref{fig:standardBS}.

\section{Numerical Calculations on a lattice}
\label{numericalSect}

\subsection{Discretization and the Relaxation Method}

We solve the self-consistent
equations of motion for the 2- and 4-point functions using a
numerical lattice method
\cite{Carrington2013}. The first step is to rotate to Euclidean space. We define the Eucledian variables:
\bea && q_0=i q_E\,,~~ \delta Z = \delta Z_E\,,~~\delta m^2=\delta m^2_E\,,~~
G=-iG_{E}\,,~~\Sigma=-i\Sigma_{E}\,, \label{euc} \\
&& \lambda = -i\lambda_E\,,~~\delta\lambda=
-i\delta\lambda_E\,,~~V=iV_{E}\,,~~ M=iM_{E}\,. \nonumber\eea 
Note that the factor $i$ in the definition of the Eucledian vertex removes the $i$ introduced in the definition of $\lambda_b$ in Eq. (\ref{lagrangian}). 
The
Dyson equation in Euclidean space is (see equation (\ref{G2defn})):
\begin{equation}
G_E^{-1}(P)=G^{-1}_{0E}(P)+\Sigma_E(P)\,.\label{Eq88}
\end{equation}
All variables from here on are Eucledian and we suppress the subscript $E$.
The coupling constant and 4-point vertex have dimension
$4-d$ so that in 2D $\lambda\sim m^2$ and in 3D $\lambda\sim m$. 

There are no ultra-divergences in the 4-point
vertices in less than four dimensions, and therefore we choose the coupling
strength counter-term to be vanishing, i.e., $\delta\lambda=0$.
For the 2-point function, the only fundamental divergences are the tadpole and sunset diagrams.
The tadpole diagram has a momentum independent divergence in 2D and is finite in 3D. The sunset diagram is finite in 2D and has a momentum independent divergence in 3D. Since we have no momentum dependent divergences, we can also set $\delta Z = 0$ and renormalize the propagator with the counter-term $\delta m^2$. To determine this counter-term we use the renormalization condition $\Sigma(0)=0$, which means we can drop the tadpole diagram.
Note that if one expands the equation of motion for the 4-vertex $V$ obtained from the 4PI effective action, or the equation for the BS vertex from the 2PI effective action, each self-energy insertion is accompanied by the mass counter-term that makes it finite, and therefore there are no sub-divergences.

\vspace*{2mm}




We use an $N^{d}$ symmetric lattice with
periodic boundary conditions. 
In 2D we use $N$ up to 16 and in 3D
$N$ up to 12. The lattice spacing is $a$ and we choose $a=2\pi/(N
m)$. In Euclidean space, each momentum component is discretized:
\begin{equation}
Q_{i}=\frac{2\pi}{aN}n_{i}=m \,n_{i}\,,\quad
n_{i}=-\frac{N}{2}+1,...,\frac{N}{2}\,.\label{momenta}
\end{equation}
Indices which fall outside of the range $\{-N/2+1,N/2\}$ are wrapped inside using periodic boundary conditions. This is done using the function
\bea
\label{reindex}
{\rm rndx}[{\rm index}] = 1 - N/2 + {\rm Mod}[{\rm index} + N/2 - 1, N]\,,
\eea
where Mod[$m$,$n$] is an integer function that gives the remainder on division of $m$ by $n$ so that $0<$ Mod[$m$,$n$] $<n-1$ (for example, Mod[17,17]=0 and Mod[23,17]=6). To illustrate we consider $N=4$. Integers outside the range $\{-1,0,1,2\}$ are wrapped into this range periodically using the function in Eq. (\ref{reindex}):
\bea
&& ~~~~~ \vdots \nonumber\\
&& \{{\rm rndx}[-5],{\rm rndx}[-4],{\rm rndx}[-3],{\rm rndx}[-2]\} \to \{-1,0,1,2\} \nonumber\\[2mm]
&& \{{\rm rndx}[-1],{\rm rndx}[0],{\rm rndx}[1],{\rm rndx}[2]\} \to \{-1,0,1,2\} \nonumber\\[2mm]
&& \{{\rm rndx}[3],{\rm rndx}[4],{\rm rndx}[5],{\rm rndx}[6]\} \to \{-1,0,1,2\} \nonumber\\
&& ~~~~~ \vdots \nonumber
\eea

We scale all dimensional quantities by the mass, or equivalently set $m=1$ and use mass units for all variables that carry dimension. Using this notation the size of the box in co-ordinate space $L=aN$ is fixed, and the maximum momentum is $p_{\rm max} = N/2$. 

On the lattice, the 4PI equations of  motion become:
\begin{eqnarray}
V(P_{a},P_{b},P_{c})&=&-\lambda  +\frac{1}{2}\frac{1}{(aN)^{d}}\sum_{Q}\Big[V(P_{a},P_{c},Q)G(Q)G(Q+P_{a}+P_{c})V(P_{b},P_{d},-Q)\nonumber \\
&&+V(P_{a},P_{b},Q)G(Q)G(Q+P_{a}+P_{b})V(P_{c},P_{d},-Q)\nonumber \\
&&+V(P_{a},P_{d},Q)G(Q)G(Q+P_{a}+P_{d})V(P_{b},P_{c},-Q)\Big]\,.\label{Eq25}
\end{eqnarray}
\begin{eqnarray}
\Sigma_{\mathrm{4PI}}(P) &=& \delta m^2 +
\frac{1}{6} \lambda
\frac{1}{(aN)^{2d}}\sum_{Q}\sum_{K}V(P,Q,K)G(Q)G(K)G(Q+K+P)\,.\label{Eq26}
\end{eqnarray}

In the 2PI formalism, we truncate the effective action at 
3-Loops. The self-energy contains the tadpole and sunset diagrams as in figure \ref{SEall}, but with all vertices bare. On the lattice we have
\begin{eqnarray}
\Sigma_{\mathrm{2PI}}(P) &=& \delta m^2 -
\frac{1}{6}\lambda^{2}
\frac{1}{(aN)^{2d}}\sum_{Q}\sum_{K}G(Q)G(K)G(Q+K+P)\,.\label{sigma}
\end{eqnarray}
The kernel $\Lambda$ contains the bare vertex and the $t$- and $u$-channel 1-loop diagrams (the third and fifth diagrams in figure \ref{f4}) with bare vertices. Abbreviating the momentum arguments $(P,-P,Q,-Q)$ as $(P,Q)$ we have
\begin{eqnarray}
\Lambda(P,Q)
 = -\lambda+\frac{\lambda^{2}}{2}\frac{1}{(aN)^{d}}\sum_{K}G(K)G(K+P-Q) + \frac{\lambda^{2}}{2}\frac{1}{(aN)^{d}}\sum_{K}G(K)G(K+P+Q),\label{lambda}
\end{eqnarray}
and the BS equation in Eq. (\ref{BSfirst-mom}) is
\begin{eqnarray}
\label{BS-discrete}
M(P,Q)
&=&\Lambda(P,Q)+\frac{1}{2}\frac{1}{(aN)^{d}}\sum_{K}\Lambda(P,K)G^{2}(K)M(K,Q)\,.
\end{eqnarray}

We use a numerical iterative method to solve the set of self-consistent equations, for both the 2PI and 4PI
theories. A naive implemenation of this method would be as follows.  

\vspace*{2mm}

\noindent 2PI:  On the right side 
of the 2PI equations (\ref{sigma})-(\ref{BS-discrete}) one starts using 
$G(P) = G_{\rm in}(P) = G_0(P)$ and $M(P,K) =M_{\rm in}(P,K)= \Lambda(P,K)$.
The first iteration of the calculation then produces 
 $G_{\rm out}(P)$ and $M_{\rm out}(P,K)$. 
 One defines
\bea
\label{update2pi}
&& G_{\rm up}(P) =\alpha G_{\rm out}(P) + (1-\alpha)G_{\rm in}(P)\,,\nonumber \\[2mm]
&& M_{\rm up} =\alpha M_{\rm out}(P,K) + (1-\alpha)M_{\rm in}(P,K)\,, 
\eea
and then uses these updated values as input for the second iteration. The parameter $1>\alpha>0$ is 
chosen to improve convergence.

\noindent 4PI:  On the right side 
of the 4PI equations (\ref{Eq25}) and (\ref{Eq26})
one starts with
$G(P) = G_{\rm in}(P) = G_0(P)$ and  $V(P,K,Q) = V_{\rm in}(P,K,Q) = -\lambda$ and first iteration of the calculation  produces 
 $G_{\rm out}(P)$ and $V_{\rm out}(P,K,Q)$. 
 The updated values are
\bea
\label{update4pi}
&& G_{\rm up}(P) =\alpha G_{\rm out}(P) + (1-\alpha)G_{\rm in}(P)\,,\nonumber \\[2mm]
&& V_{\rm up}(P,K,Q) =\alpha V_{\rm out}(P,K,Q) + (1-\alpha)V_{\rm in}(P,K,Q)\,.
\eea

The procedure we use to calculate $V$ is slightly different from what is described above.  Using the ``naive'' method would mean that after having calculated $V_{\rm out}$, but before updating, we would have values $V_{\rm out}$ and $V_{\rm in}$ at all points of the phase space.  For large values of $N$
memory constraints are an issue and therefore we use an in-place updating in which equation (\ref{update4pi}) is used at each point in phase space as the computation proceeds.  Mathematically one is iteratively
searching  for fixed points and point-wise convergence is faster using in-place updates.

The calulation is 
continued until the relative maximum difference between the input and output values at 
all points in momentum space is less than $10^{-4}$.  We use $\alpha=0.8$ and obtain convergence 
in less than 10 iterations, depending on the value of $\lambda$.  

\subsection{Time and Memory Constraints}

The number of points in the phase space of a vertex is given generically by $N^{l\times d}$ where $l$ is the 
number of independent momenta and $d$ is the dimension. For the vertex $M$ one can fix the momentum on one 
side of the vertex, because of the fact that the BS equation resumms only one channel: if one wants to 
calculate $M(P,Q_{\rm fixed})$ for any choice $Q_{\rm fixed}$ one can solve Eq. (\ref{BS-discrete}) in the form
\bea
M(P,Q_{\rm fixed})
&=&\Lambda(P,0)+\frac{1}{2}\frac{1}{(aN)^{d}}\sum_{K}\Lambda(P,K)G^{2}(K)M(K,Q_{\rm fixed})\,.
\end{eqnarray}
The phase space for the vertex $M$ is therefore $N^{1\times d}$ in spite of the fact that there are two 
independent momenta. 

The sets of self-consistent equations we are solving contain $V(P,K,Q)$, $M(P,Q_{\rm fixed})$, $G(P)$ 
and $\Sigma(P)$, but the phase space of the vertex $V$ is the largest by a huge factor, and therefore 
it is the limiting factor in terms of both memory and computing time. In this section we describe how 
one can reduce the size of the phase space of $V$ using the symmetries of the vertex. For the remainder of 
this section we specialize to three dimensions.

We represent the arguments of the 4-point function $V(P,K,Q)$ by a 3$\times$3 matrix of the form
\bea
\label{P-defn1}
{\bf P} = \left(
\begin{array}{ccc}
~p_x~ & ~p_y~ & ~p_z~ \\
~k_x~ & ~k_y~ & ~k_z~ \\
~q_x~ & ~q_y~ & ~q_z~ 
\end{array}
\right).
\eea
Since momentum is conserved the fourth leg of the vertex  has momentum $(-(p_x+k_x+q_x),
-(p_y+k_y+q_y),-(p_z+k_z+q_z))$. 
This gives us a matrix of four rows which we denote ${\bf P^\pi}$ and call the augmented matrix of  ${\bf P}$. The fourth row appears to be redundant but leads to significant gains in computational efficiency, as will be explained below. 
\bea
\label{P-defn2}
\!\!\!\!\!\!\!\!\!{\bf P^\pi} = \left(
\begin{array}{ccc}
~p_x~ & ~p_y~ & ~p_z~ \\
~k_x~ & ~k_y~ & ~k_z~ \\
~q_x~ & ~q_y~ & ~q_z~ \\
~-(p_x+k_x+q_x)~ & ~-(p_y+k_y+q_y)~ & ~-(p_z+k_z+q_z)~ 
\end{array}
\right)
 = \left(
\begin{array}{ccc}
~p_x~ & ~p_y~ & ~p_z~ \\
~k_x~ & ~k_y~ & ~k_z~ \\
~q_x~ & ~q_y~ & ~q_z~ \\
~l_x~ & ~l_y~ & ~l_z~ 
\end{array}
\right)	\,.
\eea
The values  $l_x = -(p_x+k_x+q_x),l_y=-(p_y+k_y+q_y)$,
  and $l_z=-(p_z+k_z+q_z)$ 
 are not necessarily in the range $\{-N/2+1,N/2\}$, and therefore whenever we go from ${\bf P}$ to the augmented matrix ${\bf P^\pi}$ the reindex function is invoked.

Vertices are symmetric under interchange of 
 legs and therefore any three rows of the augmented matrix ${\bf P^\pi}$  represent
 the same $V$.  The vertices have other symmetries as well: rotational symmetry
 means columns can be interchanged, and inversion symmetry allows any column to be 
 replaced by its negative.  
We want to take advantage of these symmetries to reduce computation time and memory requirements. 
The complete set of $N^9$ matrices  ${\bf P}$ can be grouped into blocks (partitions) satisfying 
$V({\bf P}_\alpha) = V({\bf P}_\beta)$
for any ${\bf P}_\alpha$ and ${\bf P}_\beta$  in the same partition.  
Two  matrices ${\bf P}$ 
and  ${\bf P'}$ are in the same partition if ${\bf P'}$ can be obtained from ${\bf P}$ by the following
procedure.
\begin{enumerate}
\item Given ${\bf P}$ form the augmented matrix ${\bf P^\pi}$ (reindexing if necessary).
\item Use any number
of the following operations in any order.
\begin{enumerate}
\item Interchange any rows of ${\bf P^\pi}$  (leg symmetry)
\item Interchange any columns of ${\bf P^\pi}$ (rotational symmetry)
\item Multiply any column(s) of ${\bf P^\pi}$ by $-1$  (inversion symmetry)
  
Reindexing is sometimes necessary because  
$-N/2\notin\{-N/2+1,N/2\}$.
\end{enumerate}
\item Choose any three rows of ${\bf P^\pi}$ to obtain ${\bf P'}$
\end{enumerate}

For example ($N = 6$),  the matrices ${\bf P}$, ${\bf P'}$ and ${\bf P''}$ in equation (\ref{example1234}) are in the same partition.
Note that the value $3$, not present in ${\bf P}$, appears in ${\bf P'}$ and ${\bf P''}$, 
which demonstrates the advantage of using the augmented matrix.
\bea
\label{example1234}
\!\!\!\!\!\!\!\!\!{\bf P} = \left(
\begin{array}{rrr}
 -2 & 0 & 1 \\
 -1 & 1 & 0 \\
 0 & 0 & -2 \\
\end{array}
\right),~
{\bf P^\pi} = \left(
\begin{array}{rrr}
 -2 & 0 & 1 \\
 -1 & 1 & 0 \\
 0 & 0 & -2 \\
3 & -1 & 1 \\ 
\end{array}
\right),~
{\bf P'} = \left(
\begin{array}{rrr}
 -1 & 1 & 0 \\
3 & -1 & 1 \\ 
 0 & 0 & -2 \\

\end{array}
\right),~
{\bf P''} = \left(
\begin{array}{rrr}
 2 & 0 & -1 \\
 1 & 1 & 0 \\
 3 & -1 & -1 \\ 
\end{array}
\right)\,.
\eea

If we define ${\bf P} \approx {\bf P'}$ whenever
${\bf P'}$ can be obtained from ${\bf P}$ using the procedure described above, then $\approx$ is an equivalence
relation on the set of all $3 \times 3$ matrices with entries in $\{ -N/2+1,N/2\}$.   
The minimal sized partition of this equivalence relation is 1, when ${\bf P}$ is the
zero matrix.  The maximal sized partitions are of cardinality $1152 = (4!)(3!)(2^3)(4)$, since 
there are $4!$ ways to interchange 
rows, $3!$ ways to interchange columns, $2^3$ ways to multiply column(s) by $-1$, and 
$4$ ways to choose three rows from a set of four.  Partitions are bigger when $N$ is
 bigger because reindexing occurs less often for large $N$.  When $N=12$ the average size of a partition is approximately
$1000$.  

We need only compute $V$ for a single element of each partition, since 
 $V({\bf P}_\alpha) = V({\bf P}_\beta)$ if ${\bf P_\alpha} \approx {\bf P_\beta}$.  
This reduces the main iteration loop by a factor of approximately the average size of a partition.  
The storage requirements for $V$ are reduced by the same factor.
The table below shows the size of the phase space and the number of representative points 
that need to be calculated for $d=3$ and different values of $N$.
\begin{center}
\begin{tabular}{|r|r|r |}  \hline
N & $N^{3\cdot (d=3)}$  & ~ smallest \# of reps~ \\\hline\hline
6 &    10,077,696                     & 11,424  \\\hline
8 &       134,217,728             &  129,502  \\\hline
10 &     1,000,000,000         &   913,661    \\\hline
12 &    ~ 5,159,780,352 ~    &    4,608,136    \\\hline
\end{tabular}
\end{center}

We choose a unique element from each partition and call it the representative of that partition, abbreviated 
 {\bf repr} or {\bf repr(P)}.  We need a list containing one representative from each partition to set up the main interative loop which calculates $V_{\rm out}({\bf repr(P)})$. 
The sums in equation (25) range over all possible ${\bf P}$ and therefore in the iterative calculation of $V_{\rm out}({\bf repr(P)})$ one needs vertices $V_{\rm in}({\bf P})$ which were not calculated in the previous iteration. 
Inside the summation we must replace $V_{\rm in}({\bf P})$ with $V_{\rm in}({\bf repr(P)})$. 
The representative-finding function must be fast because it is
the most frequently hit part of the computation and is the limiting factor in the running
time of the program.

Our first attempt was as follows. For each element ${\bf P}$, generate the 
partition containing ${\bf P}$, sort the partition, and choose its
minimal element as the representative. Save the list of representatives and use
it to determine which values of $V_{\rm out}$ to calculate. The same calculation of 
${\bf repr(P)}$ must be done over and over again on the right side of the equation for $V_{\rm out}$ inside the summation
over $V_{\rm in}$'s.  The process of generating, sorting and minimizing 1152 matrices
cannot be done $N^9$ times for large $N$.

\vskip .1in

A better method is to find a function ``${\bf vindex}$'' that computes an
integer for each ${\bf P}$ which is then used to ``address'' the value of $V_{\rm out}({\bf P}$).  The function 
 ${\bf vindex}$ should satisfy the conditions
\bea
\label{pp}
{\bf vindex(P_i) = vindex(P_j)} \;\;\; {\rm if} \;\;\; {\bf P_i \approx P_j} \,,\\[2mm]
\label{ppp}
{\bf vindex(P_i) \not= vindex(P_j)} \;\;\; {\rm if} \;\;\; {\bf P_i \not \approx P_j}\,.
\eea
Equation (\ref{pp}) says that ${\bf vindex}$ is the same for every element of a partition, and (\ref{ppp}) says that ${\bf vindex}$ is different for any two matrices which belong to different partitions.

We start by explaining how to construct the function {\bf vindex} so that (\ref{pp}) is satisfied. 
For a given augmented matrix 
${\bf P^\pi}$ the following functions are constant if rows or columns are interchanged.

\begin{itemize}
\item $f_1({\bf P^\pi}) = p_x + k_x + q_x +l_x + p_y +k_y +q_y +l_y +p_z +k_z +q_z +l_z$
\item $f_2({\bf P^\pi}) =p_x  k_x  q_x l_x  p_y k_y q_y l_y p_z k_z q_z l_z$
\item $f_3({\bf P^\pi}) =(p_x k_x q_x l_x) + (p_y k_y q_y l_y) +(p_z k_z q_z l_z)$
\item $f_4({\bf P^\pi}) =(p_x + k_x + q_x +l_x)(p_y +k_y +q_y +l_y)(p_z +k_z +q_z +l_z)$
\item $f_5({\bf P^\pi}) =(p_x + p_y + p_z)(k_x +k_y +k_z)(q_x +q_y +q_z)(l_x +l_y +l_z)$
\item $f_6({\bf P^\pi}) =p_x p_y p_z+ k_x k_y k_z+q_x q_y q_z+ l_x l_y l_z$

\end{itemize}

We define
\bea
\label{gdefn}
&& g({\bf P^\pi}) = f_1({\bf P^\pi})\cdot f_3({\bf P^\pi}) + f_5({\bf P^\pi}) +f_6({\bf P^\pi})\,,\\[2mm]
\label{nis}
&& n_i = g({\bf P^\pi })\,,~ 1 \leq i \leq 8 
~{\rm for~ 8 ~ways~ to~ multiply~ columns~ by~\{0,1,2,3\}}, -1's\,,\\
\label{vindex}
&& {\bf vindex({\bf P})} = {\rm Mod}({\rm Max} \{ n_i \} + {\rm Min} \{ n_i \} + \displaystyle \sum_{i=1}^{8} n_i, {\bf nprime})\,,
\eea
where ({\bf nprime}) is a large prime which is chosen  to correspond approximately to the available physical 
computer memory.  
The computed set  of numbers $\{ n_{i} \}$ is the same for every element of a partition and therefore {\bf vindex(${\bf P_i}$)} = {\bf vindex(${\bf P_j}$)} for any two ${\bf P_i} \approx {\bf P_j}$, which means that equation (\ref{vindex}) satisfies the condition (\ref{pp}).

However, equation (\ref{vindex}) does not always satisfy (\ref{ppp}). A
 ``collision'' occurs if ${\bf P_i} \not\approx {\bf P_j}$ but 
{\bf vindex(${\bf P_i}$)} = {\bf vindex(${\bf P_j}$)}. 
In practice, results are numerically indistinguishable whenever the
number of collisions is less than approximately $5\%$ of the total size of the set of representatives. The functions $f_i$ above, the combinations of the $f_i$'s which make $g$, and the combinations of the $n_i$'s which give ${\bf vindex}$ (equation (\ref{vindex})), are chosen so that the number of collisions is small.
We have checked that our results are not affected by collisions by
including additional functions in the set of $f_i$ and in $g$ to reduce the number of collisions. We have also run our program using a recursive definition of ${\bf vindex}$ with a different choice {\bf nprime} in the event of a collision, which reduces the number of collisions to almost zero.  
For small $N$'s one can also check the results produced using the indexing method by comparing with the original sorting/minimizing method.

\section{Numerical Results}
\label{ResultSect}

In this section we present our results.

\begin{figure}[H]
\includegraphics[scale=1.2]{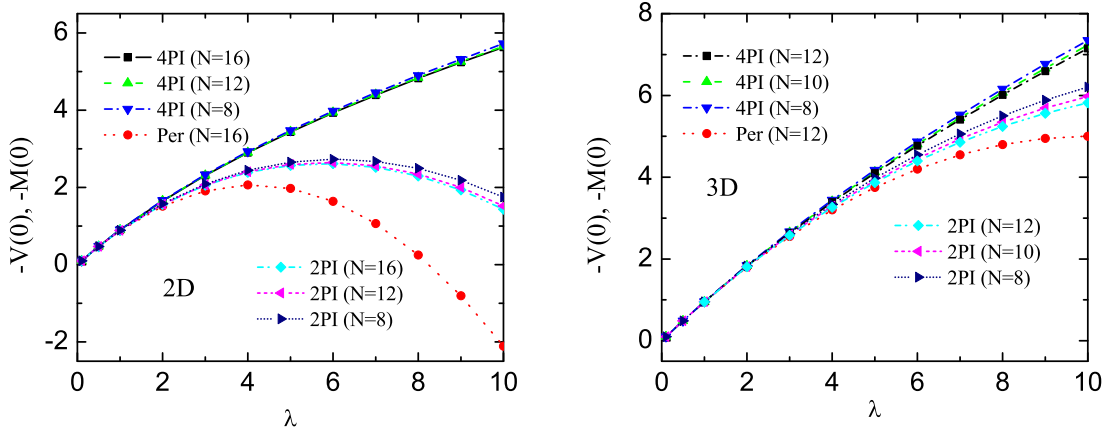}
\caption{(color online). Comparison of $V$ and $M$ as  functions of the coupling strength
$\lambda$. All the momentum arguments are chosen to be
vanishing. The calculations are done in 2D with $N=16$, 12, and 8
(left panel), and in 3D with $N=12$, 10, and 8 (right panel). For
comparison we also show the perturbative result in each graph, which
is the dotted line joining round markers (red).}\label{FigVMlam}
\end{figure}

Figure~\ref{FigVMlam} compares the 4PI self-consistent 4-vertex $V$, the 2PI BS 4-vertex $M$, and the perturbative 4-vertex, as functions
of the coupling strength. The momentum arguments are chosen to be
vanishing. The perturbative calculation is done with the
self-consistent vertex and the propagator on the right hand side of
Eq. (\ref{Eq25}) replaced by the bare ones. The 4PI and 2PI calculations
in 2D are performed with $N=16$, 12, and 8, and in 3D with $N=12$, 10, and 8. The perturbative calculation is done in 2D with $N=16$ and
in 3D with $N=12$. When the coupling $\lambda$ is small,
$V$ agrees well with $M$
and the perturbative 4-vertex, but the differences between them
become larger as $\lambda$ increases, and differences are larger in 2D than 3D. The 2D perturbative vertex becomes negative when $\lambda$ is large enough, which shows that the 1-loop contribution has overwhelmed the tree term. 
The vertex $M$ lies between $V$ and the perturbative result for all values of coupling. 

\begin{figure}[H]
\includegraphics[scale=1.2]{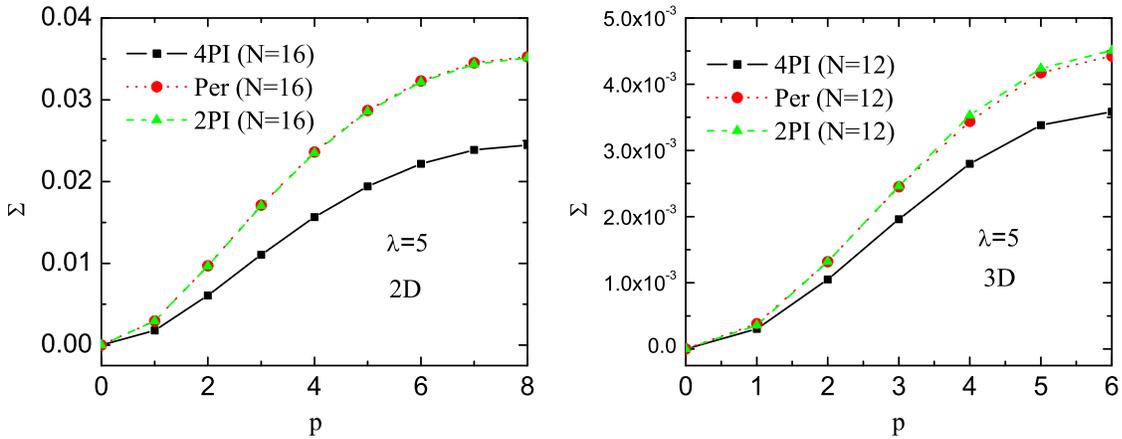}
\caption{(color online). Comparison of the dependence of the 4PI and
2PI self-energies on $P_{x}$ with all other momentum components set
to zero. Calculations are done in 2D with $N=16$ (left panel) and in
3D with $N=12$ (right panel), and $\lambda=5$. }\label{FigSigPa1}
\end{figure}

In Fig.~\ref{FigSigPa1} we look at the self-energy $\Sigma(P)$ obtained from the 4PI, 2PI, and perturbative calculations. 
We show the dependence on $P_{x}$ with all other momentum components set to zero. The 2PI self-energy agrees well with the perturbative one when $\lambda=5$ (in Ref.
\cite{Carrington2013}  it was shown that a significant difference 
between the 2PI self-energy and the perturbative one appears at very large $\lambda$). The
4PI self-energy is smaller than the 2PI and perturbative self-energies at
nonvanishing external momentum.

\begin{figure}[H]
\includegraphics[scale=1.2]{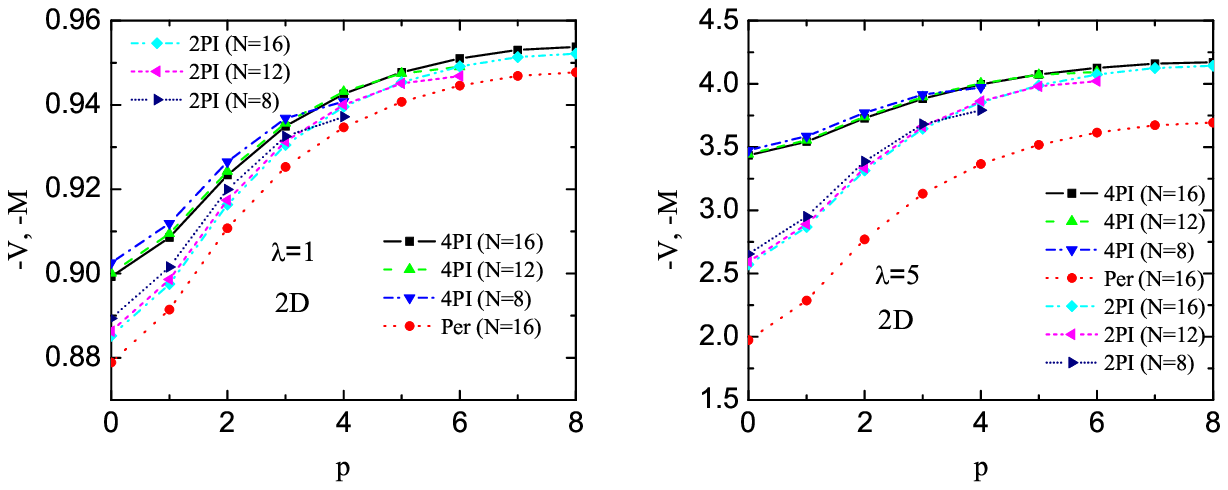}
\includegraphics[scale=1.2]{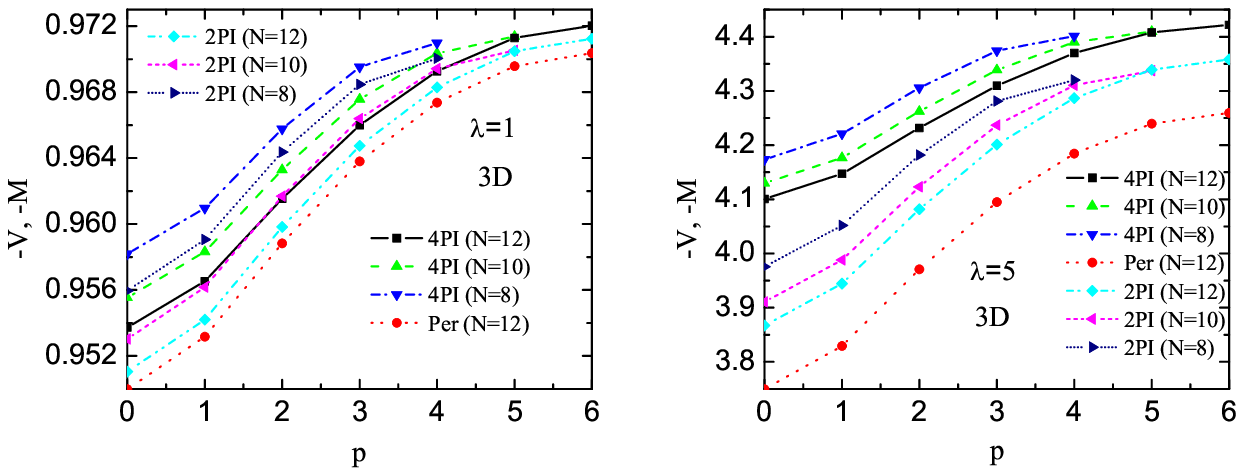}
\caption{(color online). Comparison of the dependence of $V$ and $M$ on $(P_a)_{x}$ with
all other external momentum components set to zero. Top left panel
is 2D with $\lambda=1$, top right is 2D with $\lambda=5$, bottom
left is 3D with $\lambda=1$, and bottom right is 3D with
$\lambda=5$. The perturbative result is the dotted line which joins
round markers (red).}\label{FigVMPa1}
\end{figure}

Figure~\ref{FigVMPa1} compares the dependence of $V$ and $M$ on one momentum component
$(P_a)_{x}$, with all other external momentum components set to
zero. The difference between $V$ and $M$ is
maximal at $(P_a)_{x}=0$, and decreases with increasing
$(P_a)_{x}$.
We find again that $M$ lies between $V$ and
the perturbative vertex, in this case for all momenta. The right side of the figure shows that the differences between the three vertices is larger when $\lambda$ is larger. The perturbative result deviates strongly from $V$ and $M$ at large $\lambda$, even when the external momentum is
large.

\begin{figure}[H]
\includegraphics[scale=0.7]{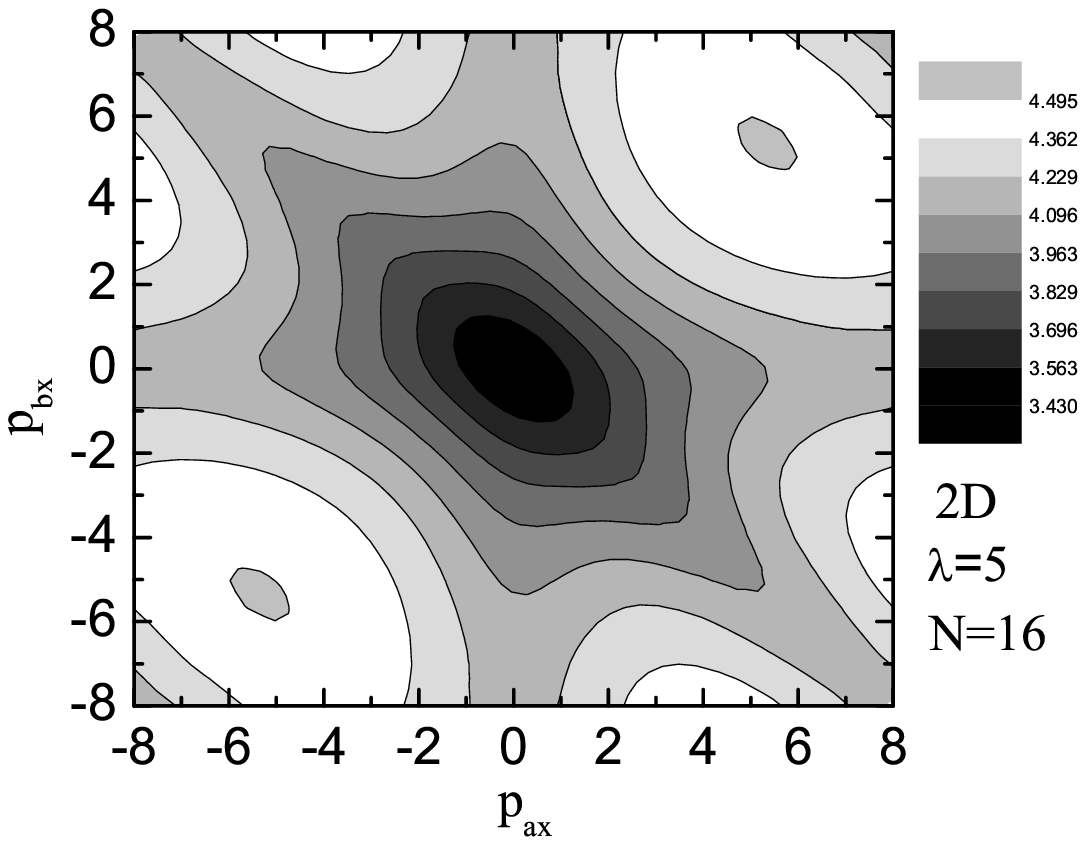}
\includegraphics[scale=0.7]{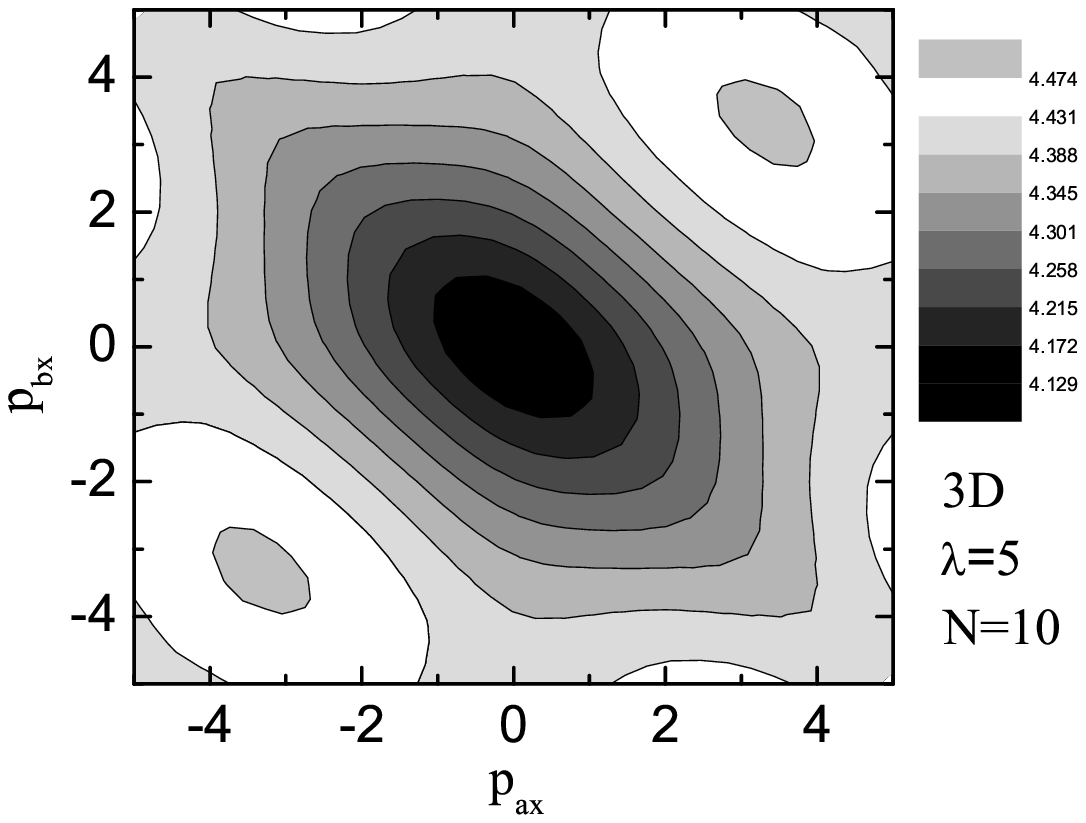}
\caption{Contour plot of $V$ in 2D and 3D as a
function of $(P_a)_x$ and $(P_b)_x$ with all other momenta
components set to zero for $\lambda=5$.}\label{f9}
\end{figure}

We also compare the vertices $V$ and $M$ for different momentum configurations. 
In Fig. \ref{f9} we show $V$ in 2D and 3D as a function of $(P_a)_x$ and $(P_b)_x$ with all other momenta components
set to zero, including all components of $P_c=0$. In 3D this can be written $(P_a;\,P_b;\,P_c) \to ((P_a)_x,0,0;\;(P_b)_x,0,0;\;0,0,0)$. 

\begin{figure}[H]
\includegraphics[scale=1.2]{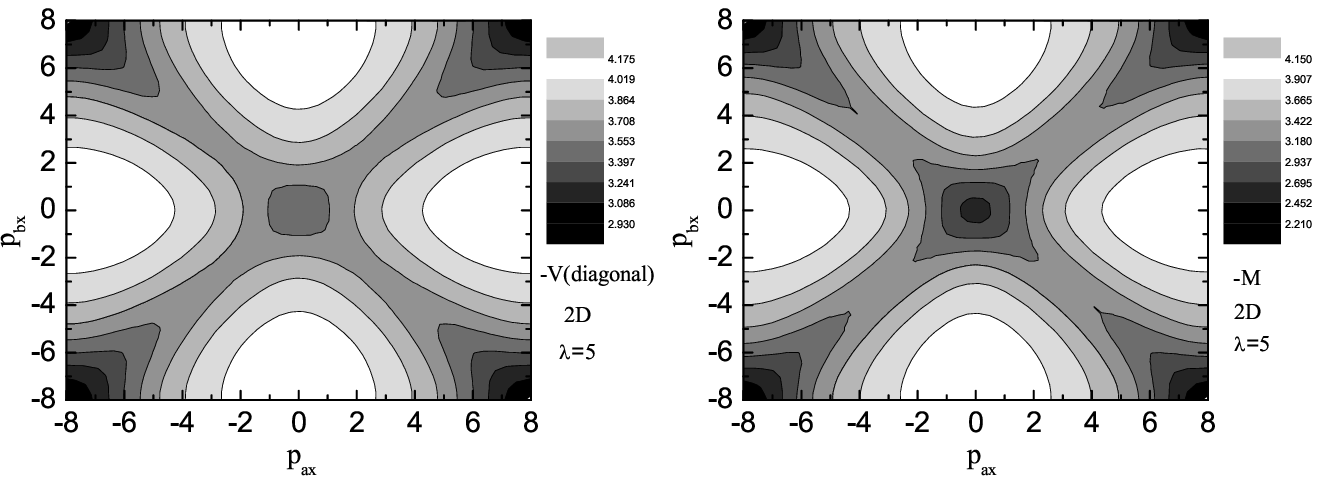}
\includegraphics[scale=1.2]{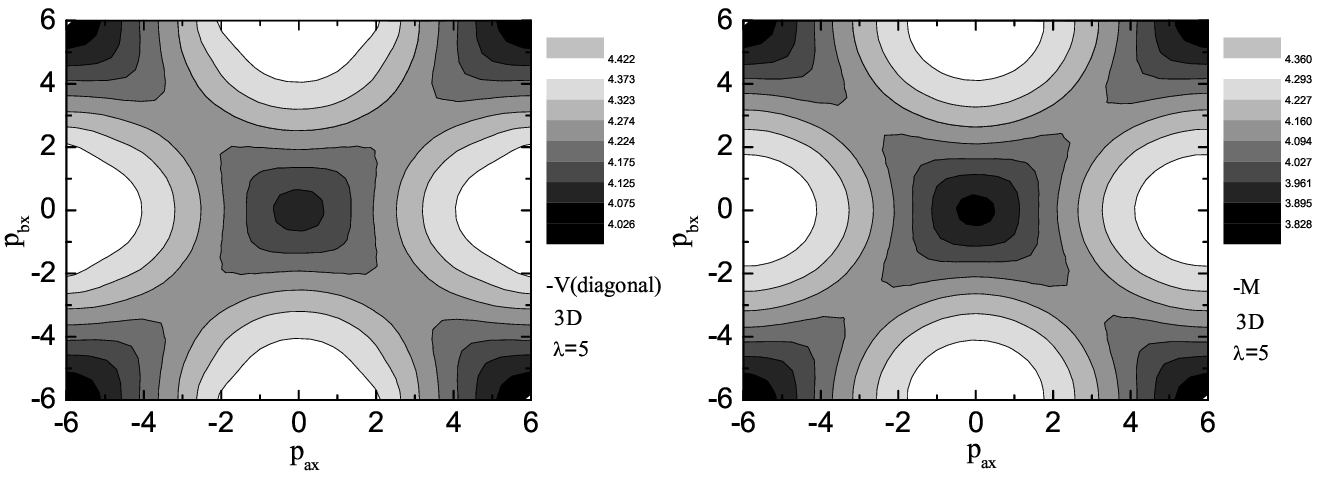}
\caption{Contour plot of 
$V((P_a)_{x},(P_b)_{x}, (P_c)_{x}=-(P_a)_{x})$ and 
$M((P_a)_{x},(P_b)_{x})$ as functions of $(P_a)_{x}$ and $(P_b)_{x}$ with
all other momentum components set to zero and $\lambda=5$. Top left panel show the
result for $V$ in 2D with $N=16$, top right for $M$ in 2D with
$N=16$, bottom left for $V$ in 3D with $N=12$, and bottom right for
$M$ in 3D with $N=12$. }\label{FigVMcont}
\end{figure}

The right side of Fig. \ref{FigVMcont} is a plot of $M((P_a)_x,0,0;-(P_b)_x,0,0)$ ($P_b$ is defined as positive coming into the vertex $V$, whereas the upper right leg of the vertex $M$ carries momentum $Q$ defined as outgoing).  Note that the curves shown in Fig. \ref{FigVMPa1} correspond to horizontal lines on the contour plots in figures \ref{f9} and the right side of figure \ref{FigVMcont}, starting at the centre of and continuing to the right edge.
 The graphs on the right side of Fig. \ref{FigVMcont} look very different from those of Fig. \ref{f9}. 
%
The left side of Fig \ref{FigVMcont} shows a contour plot of $V$ with diagonal momentum components $(P_a;\,P_b;\,P_c) \to ((P_a)_x,0,0;\;(P_b)_x,0,0;\;-(P_a)_x,0,0)$. 
Comparision of the left and right sides of Fig \ref{FigVMcont} shows that $V$ in a diagonal momentum configuration agrees well with $M$. 

\begin{figure}[H]
\includegraphics[scale=1.2]{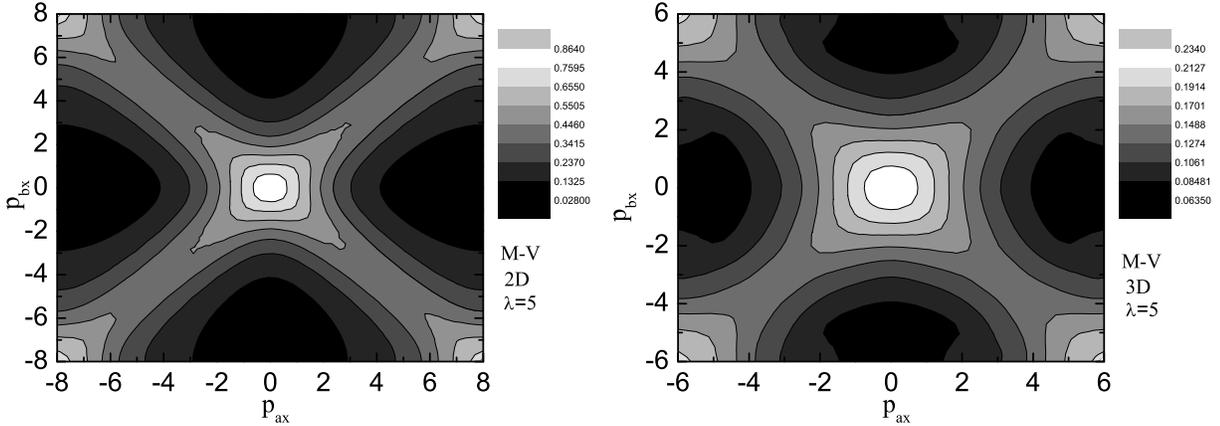}
\caption{Contour plot of $M((P_a)_{x},(P_b)_{x})-V((P_a)_x,(P_b)_x,
(P_c)_{x}=-(P_a)_{x})$ with $\lambda=5$. Calculations
are done in 2D with $N=16$ (left panel) and in 3D with $N=12$ (right
panel). }\label{FigMmV}
\end{figure}

In Fig. \ref{FigMmV} we look explicitly at the difference $$M((P_a)_x,0,0;\;-(P_b)_x,0,0)-V((P_a)_x,0,0;\;(P_b)_x,0,0;\;-(P_a)_x,0,0)\,.$$ 
The graph shows that $M-V$ is largest along the two lines which run from the upper right corner to the lower left corner, and from upper left to lower right, or the lines given by $(P_a)_x = \pm (P_b)_x$. The word 
`diagonal' has already been used to denote the configuration $P_a=P$, $P_b=-Q$, $P_c=-P$ and therefore we call the 
lines $(P_a)_x = \pm (P_b)_x$ the $\times$diagonal region. The difference $M-V$ is smallest along the horizontal and 
vertical lines that pass through the center of the graph, except for the neighbourhood of the point where these lines 
cross. Mathematically we have that $M-V$ is smallest for $(P_a)_x=0$ and $(P_b)_x\ne 0$, or $(P_b)_x=0$ and $(P_a)_x
\ne 0$. We call this the $+$diagonal region. 

We would like to understand why the difference between $V$ and $M$ is largest in the $\times{\rm diagonal}$ region and smallest in the $+{\rm diagonal}$ region. 
The vertex $M$  is a resummation in the $s$-channel of the kernel $\Lambda$, which contains the 1-loop $t$- and $u$-channels. The vertex $M$ therefore includes $s$-channels to all orders, but only the 1-loop diagrams in the $t$- and $u$- channels. The equation of motion for the vertex $V$ is a symmetric resummation in all three channels, and therefore contains $s$- $t$- and $u$-channels to all orders. 
In order to understand the relative size of $M$ and $V$ in different momentum regimes, we look at the contributions from each channel.

The three 1-loop graphs shown in Fig. \ref{f4} which are, in order, the $s$- $t$- and $u$-channels. 
For definiteness we choose one line in the $\times$diagonal and $+$diagonal regions:
\bea
&& \times{\rm diagonal}:~~~ (P_a)_x = - (P_b)_x\,,\\
&& +{\rm diagonal}:~~~ (P_a)_x \ne 0\,,~~= \pm (P_b)_x=0\,. \nonumber
\eea
In Table 1 we give the momentum arguments of the two propagators in each channel using $K$ for the loop momentum. 
\begin{center}
\begin{tabular}{|c|c|c|c |}  \hline
  & s & t & u \\\hline\hline
~ general $V(P_a,P_b,P_c,-P_a-P_b-P_c) ~ $   & ~~ $K(K+P_a+P_c)$~~ &~~ $ K(K+P_a+P_b)$ ~~ & ~~ $K(K-P_b-P_c)$ ~~ \\
diagonal $V(P_a,P_b,-P_a,-P_b)$ ~ & $K(K)$ & $K(K+P_a+P_b)$ & $K(K+P_a-P_b)$ \\
$\times$diagonal  $V(P_a,-P_a,-P_a,P_a)$ ~ & $K(K)$ & $K(K)$ & $K(K+2P_a)$ \\
$+$diagonal $V(P_a,0,-P_a,0)$ ~ & $K(K)$ & $K(K+P_a)$ & $K(K+P_a)$ \\
\hline
\end{tabular}
\end{center}

Our results indicate 
that diagrams with propagators of the form $G(K)^2$ will give a larger contribution than diagrams with propagators $G(K)G(K+K^\prime)$. This can be seen as follows. 
In the $\times$diagonal region the $s$- and $t$-channels both contain a factor $G(K)^2$ and therefore their 
contributions will be greater than that of the $u$-channel. Since $V$ contains a $t$-channel resummation and $M$ does not, we expect $M-V$ large in this region. 
In the $+$diagonal region the largest contribution comes from the $s$-channel. Since both $V$ and $M$ include an infinite resummation in the $s$-channel, we expect the difference between them to be small. 
%


\newpage

\section{Summary and Outlook}
\label{concSect}

In this work we have compared the self-consistent 4-point vertex $V$
from 4PI effective action with the BS 4-vertex $M$ from 2PI effective
action in 2D and 3D. 
In order to do the calculation we have developed a technique to fully exploit the symmetries of the 4-vertex $V$. 
The difference between $V$ and $M$ depends on the external momenta. We have shown that when $V$ is taken in a diagonal momentum configuration, the two vertices agree well when $\lambda$ is as large as 5.  
For non-diagonal momenta, 
$V$, $M$ and the perturbative vertex agree well when the coupling strength $\lambda$ is small, 
but when the coupling strength is increased the three vertices differ strongly from each other. 
We conclude that for
physical quantities where the physics of the 4-point function is
important, but for which diagonal momenta are expected to dominate,
accurate results can be expected using the 2PI effective action. In
general however, one needs the full non-perturbative 4-point
function obtained from the 4PI effective action. 

\section*{Acknowledgements}

The authors would like to thank Norm Finlay for invaluable technical assistance. 

This work was supported by the Natural and Sciences and Engineering
Research Council of Canada. WJF is supported in part by the National
Natural Science Foundation of China under Contracts No. 11005138.

\end{document}